\documentclass[11pt]{article}
\usepackage{amssymb,slashed,graphicx,color,epsf,cite}
\usepackage{mathtools}
\usepackage{stackengine}
\usepackage{tcolorbox}

\stackMath
\usepackage{blkarray}
\usepackage{amsmath}
\usepackage{comment}
\usepackage{lscape}
\def\nn{\nonumber}

\def\be{\begin{equation}}
\def\ee{\end{equation}}
\def\ben{\begin{displaymath}}
\def\een{\end{displaymath}}
\def\bea{\begin{eqnarray}}
\def\eea{\end{eqnarray}}

\makeatletter \@addtoreset{equation}{section} \makeatother

\newcommand{\w}[1]{\\[0.#1cm]}

\def\tQ{{\widetilde Q}}

\begin{document}

\begin{flushright}

\end{flushright}

\vspace{25pt}

\begin{center}

{\bf \Large {Radiatively stable ghost and tachyon freedom}\\
 \vspace{5pt} {\bf \Large {in Metric Affine Gravity}}}

\vspace{0.3in}

{\large C. Marzo$^1$}

\vspace{0.3in}

\small{\textit{$^1${\it  Laboratory of High Energy and Computational Physics, NICPB, R\"{a}vala 10, 10143 Tallinn, Estonia}}}

\vspace{40pt}

\end{center}

\vskip 0.5in

\baselineskip 16pt

\begin{abstract}{We report the existence of a ghost- and tachyon-free sector in metric-affine theories of gravity, that is invariant under diffeomorphism and a particular abelian symmetry. In contrast with many studied cases in the literature, the constraints for unitarity and causality are granted by non-accidental symmetries and do not ask for further tuning, whose fate under renormalization would be unclear. Unsurprisingly, the minimal model is massless. We find that a mechanism to provide mass is accommodated by a simple Stueckelberg extension of metric-affine gravity involving the non-metricity tensor. A non-trivial result is that also such an extension describes a ghost- and tachyon-free dynamic stabilized by the same abelian symmetry. The resulting spectrum of the collective rank-3, rank-2, and rank-0 Lagrangian is investigated with the operators recently computed in the literature.}
\end{abstract}

\vspace{15pt}

\thispagestyle{empty}

\pagebreak


\tableofcontents

\newpage

\section{Introduction}
The persistent dissatisfaction with some aspects of quantum gravity supports the search for theoretical frameworks beyond the minimal Einstein-Hilbert model. One possible avenue in this direction considers the introduction of extra fields that also transform, inhomogeneously, under diffeomorphisms. A particular realization of this idea, distinguished for its geometrical allure, is given by Metric-Affine Gravity (MAG). In this paradigm, the symmetric rank-2 tensor, which carries the graviton, is supported by a dynamical rank-3 field associated with the geometrical properties of torsion and non-metricity. As shared with similar efforts to modify gravity, such extension calls for new states which, in general, mine the underlying probabilistic and causal quantum texture. 
This fundamental problem, interconnected with the renormalization properties of gravity, is sometimes completely ignored in models that ultimately trade quantum mechanics for the solution of a given cosmological puzzle. When considered with more care, the awareness of propagating ghostly states has propelled new approaches intended to curb or eliminate their disruptive effect on unitarity and causality \cite{Salvio:2014soa,Mannheim:2006rd,Donoghue:2019fcb,Donoghue:2021eto}. An interesting development on this line has brought to light a different quantization prescription that deals with massive ghostly propagation in a way to preserve unitarity while accepting microscopic violation of causality \cite{Anselmi:2018ibi,Anselmi:2018tmf,Anselmi:2020opi}. In this particular framework, as shown in \cite{Piva:2021nyj}, asymptotic freedom can be realized in MAG. On the other side, traditional quantum field theory requires to take offence at the presence of ghosts and tachyons and to draw, accordingly, the constraints to avoid them. Indeed, the success of Yang-Mills and Einstein-Hilbert theories appear to indirectly support the relevance of ghost- and tachyon-freedom, as shown in \cite{VanNieuwenhuizen:1973fi,Gupta:1954zz,Deser:1969wk,Fang:1978rc,Deser:1963zzc}. Considering MAG, extensive literature has been produced to identify sectors of the large parameter space that allow a healthy propagation of additional particles \cite{Neville:1978bk,Neville:1979rb,Sezgin:1979zf,Karananas:2014pxa,Sezgin:1981xs,Lin:2018awc,Lin:2019ugq,Lin:2020phk,Blagojevic:2018dpz,Baikov:1992uh,Percacci:2020ddy}. The presence of high-rank fields, carriers of multiple particle states, brings to this search an intrinsic complexity\footnote{Such complexity is the trademark of using redundant fields in order to describe particles. Research is ongoing to achieve an efficient way to describe physical processes avoiding these field-related nuisances \cite{Arkani-Hamed:2017jhn,Falkowski:2020fsu,Criado:2020jkp,Criado:2021itq,Criado:2021gcb}.}. To efficiently deal with it, the use of projector operators is essential in supporting a fast, mechanical, spectral analysis. This procedure unambiguously selects the constraints on the parameters to avoid imaginary masses and negative residues at the poles of the tree-level propagator. Left out from this analysis is whether such solutions can be preserved by radiative corrections, which can reintroduce the instabilities we attempted to remove. It is interesting to recognize that in the lower rank cases of spin-1 (Yang-Mills) and spin-2 (Einstein-Hilbert), the full non-linear interactions are shaped exactly to preserve those constraints which granted ghost-freedom at the linear level.\\
In this paper, we propose to extend to MAG the same rationale. 
In this way, we discriminate among different solutions of the (tree-level) spectral analysis by dismissing those which require an unstable tuning between parameters, hence between operators. Again, the large parameter space makes such endeavour particularly challenging and the existence of possible solutions, in agreement with the mentioned principle, is highly non-trivial. We start our exploration by looking for the simplest realization of this rationale. For this purpose we rely on, as illustrated in section \ref{secSym}, stabilizing non-accidental symmetries imposed to the full MAG action \cite{Percacci:2020ddy}. Then, using an opportune classification, we can find a shortcut toward a simple and interesting minimal model. We prove that the requirement of radiative stability selects an abelian sector from MAG and predicts the presence of a vector state from the non-metricity tensor. Unsurprisingly this vector is massless, also in continuity with what we know from low-spin gauge theories. Accordingly, this points to the existence of a supportive scalar sector of  Stueckelberg/Goldstone nature. By using the extended basis of operator completed in \cite{Marzo:2021esg} we can show, also a non-trivial result, that such extension does not reintroduce any unstable tuning of parameters. \\
In this work we use the mostly minus metric signature. Given the volume of the computations, typos are inevitable. Notebooks with correct formulas can be obtained upon request to the author. 
%
\section{Metric-Affine Gravity} \label{secMAG}
%
\subsection{The Metric-Affine Gravity Action}

The target of our spectral investigation will be the so-called metric-affine theory of gravity \cite{Hehl:1994ue}  (MAG). This theory extends the dynamics of the minimal Einstein-Hilbert action by adding a rank-3 field $A_{\mu \,\,\nu}^{\,\,\,\rho}$, quadratically mixed with the dynamical symmetric tensor $h_{\mu \nu} = g_{\mu \nu} - \eta_{\mu \nu}$. The mixing, and the rest of the Lagrangian terms, are shaped by the non-linear completion of the defining gauge symmetries 
\bea \label{eq1}
&&\delta h_{\mu \nu} = \partial_{\mu} \xi_{\nu} + \partial_{\nu} \xi_{\mu} \,\, , \nn \\
&&\delta A_{\mu \,\,\, \nu}^{\,\,\,\,\rho} = \partial_{\mu} \partial_{\nu} \xi^{\rho} \,\, .
\eea
As it is well known, this collective gauge symmetry builds a theory that enjoys a powerful geometrical interpretation once the symmetric field $h_{\mu \nu}$ is integrated into the metric of curved space-time. Then eq.~(\ref{eq1}) are  completed to the familiar diffeomorphism transformation generated by $x' = x - \xi(x)$ 
\bea \label{eq2}
&& g'_{\mu \nu}(x') = \frac{\partial x^{\alpha}}{\partial {x'}^{\mu}} \frac{\partial x^{\beta}}{\partial {x'}^{\nu}} g_{\alpha \beta}(x)\, , \nn \\
&& {A'}_{\mu \,\,\, \nu}^{\,\,\,\rho}(x') = \frac{\partial x^{\alpha}}{\partial {x'}^{\mu}} \frac{\partial x^{\beta}}{\partial {x'}^{\nu}} \frac{\partial {x'}^{\rho}}{\partial x^{\sigma}} A_{\alpha \,\,\, \beta}^{\,\,\,\sigma}(x) + \frac{\partial{x'}^{\rho}}{\partial x^{\sigma}} \frac{\partial^2 x^{\sigma}}{\partial {x'}^{\mu} \partial {x'}^{\nu}} \,\,.
\eea
The shifting symmetry of $A_{\mu \,\,\, \nu}^{\,\,\,\rho}$ naturally assigns to this (Lorentz) tensor the familiar role of affine/gauge connection, so that the definitions of covariant derivative $D_{\mu}$, curvature $F_{\mu \nu \,\,\, \sigma}^{\,\,\,\,\, \rho}$ 
\bea \label{eq3}
&&D_{\mu}T_{\alpha \cdots}^{\quad \beta \cdots} = \partial_{\mu} T_{\alpha \cdots}^{\quad \beta \cdots} + A_{\mu \,\,\, \alpha}^{\,\,\, \sigma} T_{\sigma \cdots}^{\quad \beta \cdots} + \cdots - A_{\mu \,\,\, \sigma}^{\,\,\, \beta} T_{\alpha \cdots}^{\quad \sigma \cdots} + \cdots \nn \\
&& F_{\mu \nu \,\,\, \sigma}^{\,\,\,\,\, \rho} = \partial_{\mu} A_{\nu \,\,\, \sigma}^{\,\,\,\rho} + A_{\mu \,\,\, \alpha}^{\,\,\,\rho} A_{\nu \,\,\, \sigma}^{\,\,\,\alpha} - (\mu \leftrightarrow  \nu) \, , 
\eea
together with the tensors 
\bea
&& T_{\mu \,\, \nu}^{\,\,\alpha} = A_{\mu \,\,\, \nu}^{\,\,\,\rho} - A_{\nu \,\,\, \mu}^{\,\,\,\rho} \,\,, \,\,\,\,\,\,\,\,Q_{\rho \mu \nu} = - \partial_{\rho}g_{\mu \nu} + A_{\rho \,\,\, \mu}^{\,\,\,\sigma} g_{\sigma \nu} + A_{\rho \,\,\, \nu}^{\,\,\,\sigma} g_{\sigma \mu}\, ,
\eea
can be adopted in describing the invariants of the Lagrangian. $T_{\mu \,\, \nu}^{\,\,\alpha}$ and $Q_{\rho \mu \nu}$ are linked, respectively, to the geometrical properties of torsion and non-metricity, and  aid to easily distinguish different classes of theories when one, or both of them, is set to zero. \\
The resulting model, sometimes in the related form of Poincar\'{e} gauge theory, has been lengthily studied \cite{VanNieuwenhuizen:1973fi,Neville:1978bk,Neville:1979rb,Sezgin:1979zf,Karananas:2014pxa,Sezgin:1981xs,Lin:2018awc,Lin:2019ugq,Lin:2020phk,Blagojevic:2018dpz,Baikov:1992uh,Capozziello:2007tj,Gronwald:1997bx,Hehl:1999sb,Vitagliano:2010sr,Vitagliano:2013rna,BeltranJimenez:2019acz,Latorre:2017uve}. We relies, in particular, to the form of the 28 parameters MAG action presented in \cite{Percacci:2020ddy}, which we reproduce here unadulterated 
\bea \label{eq4}
&&S(g,A) = -\frac12\int d^4 x\ \sqrt{-g}\,\Big[ -a_0 F +  F^{\mu\nu\rho\sigma} \big( c_1 F_{\mu\nu\rho\sigma} + c_2 F_{\mu\nu\sigma\rho} + c_3 F_{\rho\sigma\mu\nu} + c_4 F_{\mu\rho\nu\sigma} +  
\nn\w2
&& + c_5 F_{\mu\sigma\nu\rho} + c_6 F_{\mu\sigma\rho\nu} \big) + F^{(13)\mu\nu} \big(c_7 F^{(13)}_{\mu\nu} + c_8 F^{(13)}_{\nu\mu} \big)
+ F^{(14)\mu\nu} \big( c_9 F^{(14)}_{\mu\nu} 
+ c_{10} F^{(14)}_{\nu\mu}\big) + \nn \w2  && + F^{(14)\mu\nu}\big(c_{11} F^{(13)}_{\mu\nu}
+ c_{12} F^{(13)}_{\nu\mu} \big) + F^{\mu\nu}\big(c_{13} F_{\mu\nu}
+ c_{14} F^{(13)}_{\mu\nu}
+ c_{15} F^{(14)}_{\mu\nu}\big)
+c_{16}F^2 + \nn \w2 && +  T^{\mu\rho\nu} \big(a_1 T_{\mu\rho\nu} + a_2 T_{\mu\nu\rho}\big)  + a_3 T^\mu T_\mu + Q^{\rho\mu\nu}\big( a_4 Q_{\rho\mu\nu} 
+ a_5 Q_{\nu\mu\rho}\big)  + \nn \w2 && + a_6 Q^\mu Q_\mu + a_7 \tQ^\mu \tQ_\mu + a_8 Q^\mu \tQ_\mu
 + a_9 T^{\mu\rho\nu} Q_{\mu\rho\nu}  
+ T^\mu \left( a_{10} Q_\mu 
+ a_{11}\tQ_\mu \right)
\Big]\ , \nn \\
\eea
with the contractions
\bea
 T_\mu \equiv T_\lambda{}^\lambda{}_\mu\ , \, Q_\mu \equiv  Q_{\mu\lambda}{}^\lambda\ , \, \tQ_\mu \equiv  Q_\lambda{}^\lambda{}_\mu\ , \,
F_{\mu\nu} \equiv F_{\mu\nu\lambda}{}^\lambda\ ,\, F_{\mu\nu}^{(14)} \equiv F_{\lambda\mu\nu}{}^\lambda\ , F_{\mu\nu}^{(13)} \equiv F_{\lambda\mu}{}^\lambda{}_\nu\ , \,
F \equiv F_{\mu\nu}{}^{\mu\nu}\, .\nn 
\eea
%
%
It is sometime convenient to reformulate the degrees of freedom of MAG by a redefinition of $A_{\mu \,\,\,\nu}^{\,\,\,\rho}$ that factors out the Christoffel connection $\Gamma_{\mu \nu}^{\rho}$ 
\bea \label{Atok}
A_{\mu \,\,\, \nu}^{\,\,\,\rho} = \Gamma_{\mu \nu}^{\rho} + k_{\mu \,\,\, \nu}^{\,\,\,\rho}\, . 
\eea
In this way, the MAG theory is reshaped in the form of a quadratic gravity theory augmented with $k_{\mu \,\,\, \nu}^{\,\,\,\rho}$. In terms of $k_{\mu \,\,\, \nu}^{\,\,\,\rho}$ the curvature, the torsion and the non-metricity tensor acquire the form  
\bea
&&F_{\mu \nu \,\, \sigma}^{\,\,\,\,\rho} = R_{\mu \nu \,\, \sigma}^{\,\,\,\,\,\,\rho} + \big(\nabla_{\mu}k_{\nu \,\,\sigma}^{\,\,\rho} + k_{\mu \,\,\alpha}^{\,\,\rho} k_{\nu \,\,\sigma}^{\,\,\gamma} - (\mu \leftrightarrow  \nu) \big) \, ,
\nn \w2
&&T_{\mu\,\,\nu}^{\,\,\alpha} = k_{\mu\,\,\nu}^{\,\,\alpha} - k_{\nu\,\,\mu}^{\,\,\alpha} \, , \,\, Q_{\lambda \mu \nu} = k_{\lambda \mu \nu} + k_{\lambda \nu \mu}\, ,
\eea
where the derivative operator $\nabla_{\mu}$, differently from $D_{\mu}$ in eq.~(\ref{eq3}), only includes the Christoffel connection  $\nabla_{\mu} = \partial_{\mu} + \hat{\Gamma}_{\mu}$. To distinguish between the two formulations, before and after the redefinitions (\ref{Atok}), we will refer, correspondingly, to the \emph{affine} and the \emph{contorsion} phases of the theory. 
\subsection{Propagating states in Metric-Affine Gravity} \label{alg}
A generic unconstrained quadratic Lagrangian involving the two fields $h_{\mu \nu}$ and $A_{\mu \,\,\,\nu}^{\,\,\,\rho}$ can, in general, propagate 24 particle states. These states correspond to the irreducible representation of the $SU(2)$ little group carried by the tensor representation of the Lorentz group. Each of these little group representations is identified by a spin quantum number $s$ and by one of the two eigenvalues of the parity operator $p = \pm 1$. A set of tensor fields will, in general, carry multiple representations with the same spin and parity, so a further index $j$ is conveniently introduced to enumerate them. This leads us to use the compact symbol $s^p_j$ in referring to any of the, $2 s + 1$ dimensional, $SU(2)$ representations. Using, again, the enumeration convention of \cite{Percacci:2020ddy}, we have the following decomposition for the MAG fields
\begin{eqnarray}\label{eq5}
&A_{\mu \nu \rho} \supset & 3_1^- \oplus 2_1^+ \oplus 2_2^+ \oplus 2_3^+ \oplus 2_1^- \oplus 2_2^- \oplus 1^+_1 \oplus 1^+_2 \oplus 1^+_3 \oplus 1^-_1 \oplus 1^-_2 \oplus 1^-_3 \oplus  \nn \\ && \oplus 1^-_4 \oplus 1^-_5 \oplus 1^-_6 \oplus  0^+_1 \oplus 0^+_2 \oplus 0^+_3 \oplus 0^+_4 \oplus 0^-_1 , \nn \\
&h_{\mu \nu} \supset & 2_4^+ \oplus 1^-_7 \oplus 0^+_5 \oplus 0^+_6 
\end{eqnarray}
As known, the propagation of the little group components, sourced by the derivative terms in eq.~(\ref{eq4}), will challenge the unitarity and causality of the corresponding quantum model. On a particle level, this means allowing negative norm states (ghosts) imaginary masses (tachyons) in the spectrum. \\ 
Because of the particular gauge symmetry (\ref{eq1}), it is not expected by the $h_{\mu \nu}$ to propagate anything else but a healthy $2^+$ sector
\cite{VanNieuwenhuizen:1973fi}. The same gauge symmetry is not sufficient to eliminate ghosts and tachyons once the dynamic rank-3 tensor $A_{\mu \,\,\,\nu}^{\,\,\,\rho}$ is introduced in the generic MAG theory (\ref{eq4}). This is the main problem that has propelled the search, in past and recent years, to discover and identify viable sectors in the large parameter space proper of MAG. \\
The very same issue has also helped the development of its own methodology to efficiently deal with a large parameter space and to lead, in a more direct way, to the nature of the propagating states. We will now briefly review the algorithm used while referring to \cite{Marzo:2021esg} and references therein for a more detailed exposition.  \\
The propagation of particles is determined by the quadratic part of the action $S_2$. Symbolically, we define the momentum-space kinetic term $K(q)$ to be such
\begin{equation} \label{kin}
\mathcal S_2 =  \frac{1}{2} \int d^4 q \, \bigg( \Phi(-q) \,  \mathcal \, K(q) \, \Phi(q) + \mathcal{J}(-q)\Phi(q) + \mathcal{J}(q)\Phi(-q) \bigg)\, , 
\end{equation}
where a linear coupling between fields $\Phi(q)$ and sources $\mathcal J(q)$ has been introduced. 
The goal, once $K(q)$ is known, is to compute the propagator $\mathcal D(q)$ defined via $K(q) \cdot \mathcal D(q) = \hat 1$ . Then, the poles and corresponding residues of $\mathcal D(q)$ will reveal the nature  of the propagating particles. To perform such inversion two main problems are met. The first concerns discerning the particle content, which identifies with little group elements, from the fields carrying them. The second regards the computation of $\mathcal D(q)$ in presence of gauge symmetries. Both tasks are greatly eased by the knowledge of the \emph{projector operators} $P^{i,i}_{\left\lbrace s,p\right\rbrace}$, which decompose a field into its irreducible $SU(2)$ components $s^p_i$ 
\bea
\phi_{\mu_1 \mu_2 \cdots \mu_n}\left(q\right) = \sum_{s,p,i} P^{i,i}_{\left\lbrace s,p\right\rbrace}{}_{\mu_1 \mu_2 \cdots \mu_n}^{\quad\quad \nu_1 \nu_2 \cdots \nu_n}\left(q\right) \phi_{\nu_1 \nu_2 \cdots \nu_n}\left(q\right) \,,
\eea 
and the \emph{mixing operators}, $P^{i,j}_{\left\lbrace s,p\right\rbrace}$ with $i\neq j$, which connect the different sectors $s^p_i$ and $s^p_j$ with same value of spin and parity. 
Such sets, which with an abuse of nomenclature we will collectively refer to simply as \emph{projector operators}, satisfy the completeness, orthogonality and hermitianicity relations   
\bea \label{algebraP}
&&\sum_{s,p,i} P^{i,i}_{\left\lbrace s,p\right\rbrace}{}_{\mu_1 \mu_2 \cdots \mu_n}^{\quad\quad \nu_1 \nu_2 \cdots \nu_n} = \displaystyle{\hat 1}{}_{\mu_1 \mu_2 \cdots \mu_n}^{\quad\quad \nu_1 \nu_2 \cdots \nu_n} , \nn \\ &&\nn \\
&& P^{i,k}_{\left\lbrace s,p\right\rbrace}{}_{\mu_1 \mu_2 \cdots \mu_n}^{\quad\quad \rho_1 \rho_2 \cdots \rho_n} \,\, P^{j,w}_{\left\lbrace r,m\right\rbrace}{}_{\rho_1 \rho_2 \cdots \rho_n}^{\quad\quad \nu_1 \nu_2 \cdots \nu_n} = \delta_{k, j}\, \delta_{s, r}\, \delta_{p, m}\, P^{i,w}_{\left\lbrace s,p\right\rbrace}{}_{\mu_1 \mu_2 \cdots \mu_n}^{\quad\quad \nu_1 \nu_2 \cdots \nu_n} , \nn \\ &&\nn \\
&& P^{i,j}_{\left\lbrace s,p\right\rbrace}{}_{\mu_1 \mu_2 \cdots \mu_n}^{\quad\quad \nu_1 \nu_2 \cdots \nu_n} = 
\left(P^{j,i}_{\left\lbrace s,p\right\rbrace}{}^{\nu_1 \nu_2 \cdots \nu_n}_{\quad \quad \mu_1 \mu_2 \cdots \mu_n}\right)^*  .
\eea
The role of the projector operators is difficult to overestimate and a relevant part of the research of ghost- and tachyon-free theories is dedicated to explicitly solving the algebra (\ref{algebraP}) in order to find them. For the applications strictly relevant to first-order gravity only recently the full set of relevant operators has been computed \cite{Percacci:2020ddy}. 
Knowing the form of the $P^{i,j}_{\left\lbrace s,p\right\rbrace}$ it is possible to rewrite the kinetic term in (\ref{kin}) as 
\begin{eqnarray}
&&\int d^4 q \,  \Phi(-q) \, K(q) \, \Phi(q) \, =  \int d^4 q \, \Phi(-q) \, \sum_{s,p,i,j} \bigg( a^{\left\lbrace S,p \right\rbrace}_{i,j}  P^{i,j}_{\left\lbrace s,p \right\rbrace} \bigg) \, \Phi(q)  \,\,, 
\end{eqnarray}
where  
\bea \label{eq9}
a_{i,j}^{\left\lbrace s,p \right\rbrace} = \frac{1}{2 s - 1} \,\, Tr\, P^{i,j}_{\left\lbrace s,p \right\rbrace} \, \mathcal K (q) \,,
\eea
and the trace is over the hidden Lorentz indices \cite{Marzo:2021esg}. The projector operators trade, therefore, the complexity of the index structure for the more simple set of matrices $a^{\left\lbrace s,p \right\rbrace}_{i,j}$. 
The central problem of the inversion of the kinetic terms in order to get the propagator $\mathcal D(q)$ 
\bea \label{eq10}
&& K(q) \, \mathcal D(q) \, =   \sum_{S,p,i,j} \bigg( a^{\left\lbrace S,p \right\rbrace}_{i,j}  P^{i,j}_{\left\lbrace S,p \right\rbrace} \bigg) \, \mathcal D(q) =  \hat{1} \,\,, 
\eea
can now be solved once the inverse matrices $b^{\left\lbrace s,p \right\rbrace}_{i,j} = \left(a^{\left\lbrace s,p \right\rbrace}_{i,j}\right)^{-1}$ are computed,  
\bea \label{eq10b}
\mathcal D(q) = \sum_{s,p,i,j}  b^{\left\lbrace s,p \right\rbrace}_{i,j}  P^{i,j}_{\left\lbrace s,p \right\rbrace} \,\,.
\eea 
When such inversion is not possible the model displays gauge symmetries directly connected with the $n$ null-vectors $X_i^{r = 1,2\cdots n}$ of the matrix $a^{\left\lbrace s,p \right\rbrace}_{i,j}$ 
\bea \label{eq11}
\delta \Phi = X^r_i \, P^{i,j}_{\left\lbrace s,p \right\rbrace} \Psi \,, \quad\,\,(r = 1,2\cdots n)\,,
\eea
where $\Psi$ is an arbitrary, momentum-dependent gauge parameter. Again, beware of our concise notations where Lorentz contraction are hidden and symbols in capital Greek letters represents multiple fields \cite{Marzo:2021esg}.  
To find the form of $\mathcal D (q)$, a gauge fixing is necessary which, in this formalism, is achieved by selecting a non-degenerate submatrix $\tilde{a}^{\left\lbrace s,p \right\rbrace}_{i,j}$. This matrix can be inverted with no obstruction defining $\tilde{b}^{\left\lbrace s,p \right\rbrace}_{i,j} = \left(\tilde{a}^{\left\lbrace s,p \right\rbrace}_{i,j}\right)^{-1}$. 
The arbitrariness in the choice of the invertible submatrix reflects the freedom of gauge fixing. To counterbalance it, we base our spectral assessments on the \emph{saturated propagator} $\mathcal D_S(q)$ \cite{VanNieuwenhuizen:1973fi}
\bea \label{eq13}
\mathcal D_S(q) =  \tilde{J}^*\left(q\right) \bigg(\sum_{s,p,i,j}  \tilde{b}^{\left\lbrace s,p \right\rbrace}_{i,j}  P^{i,j}_{\left\lbrace s,p \right\rbrace}\bigg)  \tilde{J}\left(q\right) 
\eea 
which is gauge invariant provided that 
\bea \label{eq12}
X^{*}{}^r_j \,\, P^{i,j}_{\left\lbrace s,p \right\rbrace} \,\, \tilde{J} \left(q\right) = 0 \,, \quad\,\,(r = 1,2\cdots n)\,\, .
\eea
The saturated propagator is the last step to determine unambiguously whether the quadratic Lagrangian supports ghostly or tachyonic states. To compute it, our algorithm will reformulate eq.(\ref{eq13}) and (\ref{eq12}) in components, for an opportune frame of reference, thus reducing the tensor equations to simple algebraic ones. 
\subsection{Affine vs Contorsion phase, a lesson from Neville's model}
%
As a paradigmatic case to illustrate some unsatisfactory facets of seeking non-pathological propagating sectors in the large space of MAG, and to set some definitions for future use, we reanalyze the emergence, from metric MAG, of the Neville model \cite{Neville:1978bk}. Such model adds to the graviton a massive scalar from the $0^-$ sector conveyed by $A_{\mu \,\,\,\nu}^{\,\,\,\rho}$. In the affine phase, the Neville model is a particular type of the 10-parameters metric MAG theories ($Q_{\mu \nu \rho}=0$). Once again we use the formulas of \cite{Percacci:2020ddy}
\bea \label{MetricAction}
S(g,A) &=&-\frac12
\int d^4 x\ \sqrt{-g}\,\Big[-a_0 F+ F^{\mu\nu\rho\sigma} 
\big(g_1 F_{\mu\nu\rho\sigma} 
+g_3 F_{\rho\sigma\mu\nu}
+g_4 F_{\mu\rho\nu\sigma}) + \nn
\w2
&&  
+ \, F^{(13)\mu\nu} \big(g_7 F^{(13)}_{\mu\nu} 
+g_8 F^{(13)}_{\nu\mu} \big)
+ g_{16} F^2
+ T^{\mu\rho\nu} \big(b_1 T_{\mu\rho\nu} 
+b_2 T_{\mu\nu\rho}\big)  
+ b_3 T^\mu T_\mu
\Big]\ . \nn \\
\eea
The usual path to polish the parameter space is to compute the spin/parity matrices $a^{\left\lbrace s,p \right\rbrace}_{i,j}$, find the corresponding saturated propagator, and gather the forms, in terms of the couplings, of masses and residues. These formulas are then scrutinized in order to find those coherent values for the couplings leading to all positive masses and residues. In doing so, constraints are imposed on the theory's parameters which, in general, will not be preserved by the effective quantum action. 
Indeed, to recover the Neville model, the simplifying assumption $g_7 = g_8 = g_{16} = b_1 = b_2 = b_3 =0$ is assumed to start with. \\
When redefining the affine connection $A_{\mu \,\,\,\nu}^{\,\,\,\rho}$ as in (\ref{Atok}), the action (\ref{Nev1}) will generate a peculiar form of a quadratic gravity theory with a structure that we symbolically write as
\bea \label{phase12}
S(g,k) = \int d^4 x\ \sqrt{-g}\,\Big[\mathcal L (\mathcal R, \mathcal R^2) + \mathcal L(\mathcal R,  k) + \mathcal L( k)\Big] \,.  
\eea
In (\ref{phase12}) the symbol $\mathcal L (\mathcal R, \mathcal R^2)$ refers to Lagrangian terms which are a functional of Riemann curvature only, meanwhile, $\mathcal L (\mathcal R, k)$ describes mixed Riemann-contorsion terms, possibly including covariant derivative terms with powers of $\nabla_{\mu}$. Lastly, with $\mathcal L(k)$, we symbolize the possible polynomial terms up to fourth-power in the contorsion field.
Notice that in using (\ref{Atok}) we have automatically imposed the metric condition $Q_{\mu \nu \rho}=0$ by working with an antisymmetric $k_{\mu \nu \rho} = - k_{\mu \rho \nu }$, therefore the states are 
\begin{eqnarray}\label{Nev1}
&k_{\mu \nu \rho} \supset &  2_3^+ \oplus 2_2^- \oplus  1^+_2 \oplus 1^+_3 \oplus  1^-_3  \oplus 1^-_6 \oplus  0^+_3  \oplus 0^-_1 , \nn \\
&h_{\mu \nu} \supset & 2_4^+ \oplus 1^-_7 \oplus 0^+_5 \oplus 0^+_6 \,.
\end{eqnarray}
When operating the transition to the contorsion phase two distinct scenarios emerge corresponding to the presence, or absence, of mixing between Riemann curvature $R_{\mu\,\,\nu \rho}^{\,\,\sigma}$ and the contorsion field $k_{\mu \nu \rho}$. In absence of $\mathcal L (\mathcal R, k)$, which we refer to as the \emph{decoupled} scenario, the curvature squared operators in $\mathcal L (\mathcal R, \mathcal R^2)$ signals, with few exceptions\footnote{See, for instance, related references and discussion in \cite{Sezgin:1979zf}.}, the presence of ghosts in the spectrum. It is, therefore, a reasonable and safe  assumption, in this particular case, to set to zero the combination of parameters that, in the contorsion phase, multiply the independent \footnote{Different quadratic contractions of the Riemann tensor are connected, in four dimensions, by a total derivative.} curvature squared terms. \\
Conversely, in the \emph{coupled} scenario, the presence of $\mathcal R^2$ is not, automatically, a token for ghost states. In this case, as we check explicitly, the terms in $\mathcal L (\mathcal R, k)$ might rescue the unitarity through their interplay with purely curvature terms.  \\
For the restricted model (\ref{MetricAction}) with all but $a_0, g_1, g_3$ and $g_4$ non-zero, the transition to the contorsion phase generates a quadratic gravity framework with
\bea
&&\mathcal{L}(\mathcal R, \mathcal R^2) = \tfrac{1}{2} a_0  R - \tfrac{1}{4} \left(2 g_1 + 2 g_3 + g_4 \right) R_{\mu \nu \rho \sigma} R^{\mu \nu \rho \sigma} \,, \w2
&&\mathcal{L}(\mathcal R, k) = 
 2 (g_1 + g_3) k^{\alpha \beta \nu } k^{\mu }{}_{\beta }{}^{\rho } R_{\alpha \mu \nu \rho } + g_4 k^{\alpha \beta \nu } k^{\mu }{}_{\beta }{}^{\rho } \
R_{\alpha \nu \mu \rho } + \nn \w2 && + g_4 k^{\alpha \beta \nu } k^{\mu }{}_{\beta }{}^{\rho } R_{\alpha \rho \nu \mu} - 2 g_4 R_{\alpha \beta \nu \mu } \nabla^{\mu }k^{\alpha \beta \nu } + 2 (g_1 + g_3) R_{\alpha \mu \beta \nu } \nabla^{\mu }k^{\alpha \beta \nu} \,,
\eea
plus the intimidating
\bea
&&\mathcal{L}(k) = \nn \w2
&&\tfrac{1}{2} a_0\,k^{\alpha \beta \nu } k_{\beta \alpha \nu } + \tfrac{1}{2} a_0\, k^{\alpha }{}_{\alpha }{}^{\beta } k^{\nu }{}_{\beta \nu } + \tfrac{1}{2} g_4\, k_{\alpha }{}^{\mu \rho } k^{\alpha \beta \nu} k_{\beta \mu }{}^{\sigma } k_{\rho \nu \sigma } -  g_1\, k_{\alpha \beta }{}^{\mu } k^{\alpha \beta \nu } k_{\rho \mu \sigma } k^{\rho}{}_{\nu }{}^{\sigma } + \nn \w2  && -  g_3\, k^{\alpha \beta \nu } k_{\beta }{}^{\mu \rho } k_{\mu \nu }{}^{\sigma } k_{\sigma \alpha \rho }  -  \tfrac{1}{2} g_4\, k_{\alpha \beta }{}^{\mu } k^{\alpha \beta \nu } k^{\rho }{}_{\nu }{}^{\sigma } k_{\sigma \mu \rho } -  g_3\, k^{\alpha \beta \nu } k_{\beta \alpha }{}^{\mu } k^{\rho }{}_{\nu }{}^{\sigma } k_{\sigma \mu \rho } -  g_4\, k_{\alpha }{}^{\mu \rho } k^{\alpha \beta \nu } k_{\beta \mu }{}^{\sigma } k_{\sigma \nu \rho } +\nn  \w2 && + g_1\, k_{\alpha }{}^{\mu \rho } k^{\alpha \beta \nu } k_{\sigma \nu \rho } k^{\sigma}{}_{\beta \mu } + g_4\, k^{\alpha \beta \nu } k^{\mu }{}_{\beta }{}^{\rho } \nabla_{\alpha }k_{\nu \mu \rho }  - 4 g_1\, k^{\alpha \beta \nu } k^{\mu}{}_{\beta }{}^{\rho } \nabla_{\mu }k_{\alpha \nu \rho } + \nn  \w2 && -  g_4\, k^{\alpha \beta \nu } k^{\mu }{}_{\beta }{}^{\rho } \nabla_{\mu }k_{\nu \alpha \rho } -  g_4\, \nabla_{\alpha }k_{\beta \nu \mu } \nabla^{\mu}k^{\alpha \beta \nu } + g_1\, \nabla_{\alpha }k_{\mu \beta \nu } \nabla^{\mu }k^{\alpha \beta \nu } -  g_1\, \nabla_{\mu }k_{\alpha \beta \nu } \nabla^{\mu }k^{\alpha \beta \nu }  + \nn  \w2 && - \tfrac{1}{2} g_4\, \nabla_{\mu }k_{\beta \alpha \nu } \nabla^{\mu }k^{\alpha \beta \nu } -  \tfrac{1}{2} g_4\, \nabla^{\mu }k^{\alpha \beta \nu } \nabla_{\nu}k_{\alpha \beta \mu } -  g_4\, k^{\alpha \beta \nu } k^{\mu }{}_{\beta }{}^{\rho } \nabla_{\nu }k_{\alpha \mu \rho } - 2 g_3\, \nabla^{\mu }k^{\alpha \beta \nu } \nabla_{\nu }k_{\beta \alpha \mu } + \nn  \w2 && -  g_4\, k^{\alpha \beta \nu } k^{\mu }{}_{\beta }{}^{\rho } \nabla_{\rho }k_{\alpha \nu \mu } - 4 g_3\, k^{\alpha \beta \nu } k^{\mu }{}_{\beta}{}^{\rho } \nabla_{\rho }k_{\nu \alpha \mu } \,\,,
\eea
where we neglected total derivatives. \\
Both phases are, when coming to on-shell spectral properties, obviously equivalent. An explicit comparison is easy to perform and provides a consistency check of the algorithm outlined in sec.~[\ref{alg}]. While we carried our analysis in both phases, for concision, we will only illustrate the computation in the contorsion phase.\\
With help of eq.~(\ref{eq9}), the spin/parity matrices for the sources of the fields $h_{\mu \nu}$ and $k_{\mu \nu \rho}$ can be derived. Considering the content  eq.~(\ref{Nev1}), we have the list
\bea 
&& a^{\left\lbrace 2,+ \right\rbrace}_{i,j} = 
\begin{pmatrix}
a_{3,3} & a_{3,4} \\
a_{4,3} & a_{4,4} 
\end{pmatrix}
=
\begin{pmatrix}
\tfrac{1}{2}\left(a_0 - 2 (2 g_1 + 2 g_3 + g_4) q^2\right) & \tfrac{i}{\sqrt{2 q^2}}(2 g_1 + 2 g_3 + g_4)q^4  \\
-\tfrac{i}{\sqrt{2 q^2}}(2 g_1 + 2 g_3 + g_4)q^4 & -\tfrac{1}{4} q^2 \left(a_0 + 2 (2 g_1 + 2 g_3 + g_4) q^2\right)
\end{pmatrix} \, ,  \nn \w2 
&&a^{\left\lbrace 2,- \right\rbrace}_{2,2}  = \tfrac{1}{2}\left(a_0 - (4 g_1 + g_4)q^2 \right) \,, \nn \w2 
&& a^{\left\lbrace 1,+ \right\rbrace}_{i,j} = 
\begin{pmatrix}
a_{2,2} & a_{2,3} \\
a_{3,2} & a_{3,3} 
\end{pmatrix}
=
\begin{pmatrix}
\tfrac{1}{6}\left(3 a_0 + 4 (g_3 - g_1)q^2\right) & \tfrac{2}{3}\sqrt{2}(g_3 - g_1)q^2  \\
\tfrac{2}{3}\sqrt{2}(g_3 - g_1)q^2 & -a_0 + \tfrac{4}{3}q^2(g_3-g_1) 
\end{pmatrix} \,, \nn \w2 
&& a^{\left\lbrace 1,- \right\rbrace}_{i,j} = 
\begin{pmatrix}
a_{3,3} & a_{3,6} & a_{3,7} \\
a_{6,3} & a_{6,6} & a_{6,7} \\
a_{7,3} & a_{7,6} & a_{7,7}
\end{pmatrix}
=
\begin{pmatrix}
\tfrac{1}{2}\left(-a_0 - (4 g_1 + g_4) q^2\right) & - \tfrac{a_0}{\sqrt{2}} & 0  \\
-\tfrac{a_0}{\sqrt{2}} & 0 & 0 \\
0 & 0 & 0 
\end{pmatrix} \, ,\nn \w2 
&& a^{\left\lbrace 0,+ \right\rbrace}_{i,j} = 
\begin{pmatrix}
a_{3,3} & a_{3,5} & a_{3,6} \\
a_{5,3} & a_{5,5} & a_{5,6} \\
a_{6,3} & a_{6,5} & a_{6,6}
\end{pmatrix}
=
\begin{pmatrix}
-a_0 - (2 g_1 + 2 g_3 + g_4) q^2 & \tfrac{i}{\sqrt{2 q^2}} (2 g_1 + 2 g_3 + g_4) q^4 & 0  \\
- \tfrac{i}{\sqrt{2 q^2}} (2 g_1 + 2 g_3 + g_4) q^4 & \tfrac{1}{2} q^2 (a_0 - (2 g_1+ 2 g_3 + g_4)q^2) & 0 \\
0 & 0 & 0 
\end{pmatrix} \,, \nn \w2 
&& a^{\left\lbrace 0,- \right\rbrace}_{1,1}  = a_0 + (g_4 - 2 g_1)q^2 \,.
\eea
In general, as can be inferred by the rank of the spin/parity matrices, only the sector $1^-$ and $0^+$ are degenerate. This is the known realization of gravitational diffeomorphism invariance. We now present the results of the spectral analysis obtained by following the procedure outlined in \cite{Marzo:2021esg}. Beyond the massless pole, two massive ones are sourced by the $2^-$ sector, $m^2_{2^-} = \tfrac{a_0}{g_4 + 4 g_1}$, and the $0^-$ one, $m^2_{0^-} = \tfrac{a_0}{g_4 - 2 g_1}$ . For all three we must study the form of the saturated propagator $\mathcal D_S (q^2) = \sum_{s,p} \mathcal D_S^{s,p} (q^2)$, taking into account (gravitational) gauge invariance, so to infer the sign of the residue. We found the structures
\bea
\lim_{q^2 \rightarrow m^2_{2^-}} \mathcal D_S = \frac{1}{q^2 - m^2_{2^-}}X^{\dagger} \mathcal M \,X = \frac{1}{q^2 - m^2_{2^-}} \left(\frac{4}{4 g_1 + g_4}\right)\sum_{i = 1,5}|S_i|^2 \, ,
\eea
and 
\bea
\lim_{q^2 \rightarrow m^2_{0^-}} \mathcal D_S = \frac{1}{q^2 - m^2_{0^-}}X^{\dagger} \mathcal M \,X = \frac{1}{q^2 - m^2_{0^-}} \left(\frac{2}{2 g_1 - g_4}\right)|S_0|^2 \, ,
\eea
where we use the notation $S_i$ to refer to a particular linear combination of the source components. 
The nature of the corresponding propagating particles is controlled by the sign of $a_0$ which, in turn, is connected with the properties of the massless $2^+$ sector. 
The massless limit of the saturated propagator is found of the form 
\bea
\lim_{q^2 \rightarrow 0} \mathcal D_S = \frac{1}{q^2}X^{\dagger} \mathcal M \,X = \frac{1}{q^2} \left(\frac{2}{-a_0}\right)\sum_{i = 1,2}|S_i|^2 \, .
\eea
It is therefore straightforward to conclude that a physical gravitational propagation, which requires $a_0 < 0$, would always imply a tachyonic $2^-$ state.
This propagation can be nullified by requiring $g_4 = - 4 g_1$, a constraints that does not introduce extra gauge symmetries, leaving a healthy $0^-$ particle of mass $m^2_{0^-} = {-a_0}/{6 g_1}$ in the spectrum. \\
The exploration of the simple Neville`s model answers many of the questions that are met when seeking for a unitary and causal parameter space. By studying the model in both phases it is possible to draw a clearer image of how the dipole ghosts of quadratic gravity are carried by the action in the affine phase. Moreover, their cancellation from the spectrum by setting to zero the quadratic Riemann terms in the contorsion phase is only effective in the decoupled scenario. We have shown that the requirement $g_3 = 0$ in this model is not required for this particular purpose. Finally, we stress that all the numerous constraints we imposed to the metric MAG eq.~(\ref{MetricAction}) in order to get to the final Neville theory are not, in any way, expected to be preserved by loop corrections. This concerns also the relations obtained to forbid ghosts and tachyons.   
%
\section{Use of Symmetries in MAG} \label{secSym}
%
To preserve the structure of the relevant and marginal operators which define our starting Lagrangian, a call for symmetries is mandatory. The presence of a symmetry not only will restrict the shape of the admissible operators, possibly introducing reduction relations between different couplings, but will also protect the established unitarity from radiative corrections. Our aim is to apply to MAG the same program which, in a bottom-up approach, it is possible to discern from the success of Yang-Mills and Einstein theories. As established in \cite{VanNieuwenhuizen:1973fi,Gupta:1954zz,Deser:1969wk,Fang:1978rc,Deser:1963zzc} it is indeed possible to understand the structure of such theories as the unique non-linear realizations of the emergent symmetries which rid the propagator of ghostly states. 
In light of this, a possible avenue is to adopt a rather radical approach to build radiatively stable unitary and causal models. This would entail starting with the most generic Lorentz invariant Lagrangian (at a fixed polynomial order), scan over all the possible gauge symmetries, and select those setups which provide unitarity and causality without further parameter tuning, aside from sign adjustments. While we see this as a promising, albeit challenging, path towards possible UV-complete extension of gravity, this work will focus on the narrower target of MAG theory. This means that, on top of (\ref{eq1}), we look for non-accidental symmetries \cite{Percacci:2020ddy} which are exhibited by a subspace of the MAG-type Lagrangians (\ref{eq4}).  
A relevant example is given by the \emph{projective gauge symmetry} 
\bea \label{proj}
&& \delta h_{\mu \nu} = 0 \, ,\nn \w2 
&& \delta A_{\mu \,\, \nu}^{\,\,\rho}  =  X_1 \, g_{\mu \nu} \, \xi^{\rho} + X_2 \, \delta_{\mu}^{\rho} \, \xi_{\nu} + X_3 \, \delta_{\nu}^{\rho} \, \xi_{\mu} \, ,
\eea 
where the couplings $X_1, X_2$ and $X_3$ are fixed based on the symmetry property of $A_{\mu \,\, \nu}^{\,\,\rho} $ and $\xi^{\mu}(x)$ is the local parameter.
The consequences of the invariance under (\ref{proj}) have been widely investigated \cite{Iosifidis:2019fsh,Aoki:2018lwx,Kalmykov:1994fm,Kalmykov:1995ab,Percacci:2020ddy}. We highlight, in particular, how their role to guarantee a theory free of Ostrogradsky  ghosts was analyzed in \cite{Aoki:2019rvi} for the more limited case of linear curvature theories. 
How can we proceed in pinpointing a non-accidental symmetry with chances of automatically provide a ghost- and tachyon-free MAG theory? Again, a long brute two-step scan, generating the symmetry and analyzing the spectrum, might bring to light some interesting solutions. It is our ambition to ultimately undertake such large exploration, but our previous analysis of the Neville model points to a more natural shortcut, based on the link, visible in the contorsion phase, between higher-derivative ghosts and the mixed curvature/contorsion terms. Simply put, we will only proceed to the full spectral analysis for those symmetric MAG theories that, upon switching to the contorsion phase, will obey the following requirements\\
\begin{tcolorbox} 
I) \,\,$\mathcal L (R,k) = 0$.  \\ 
II) The quadratic terms in the Riemann curvature are automatically absent.  
\bea \label{TableGray}
\eea
\end{tcolorbox}
The rationale (\ref{TableGray}) simplifies our quest by limiting the access to the time-consuming spectral analysis to a very narrow set of symmetries. We remind that (\ref{TableGray}) is by no mean a guarantee of ghost-freedom, being only a subset of ghosts of Ostrogradsky type, and a full analysis must be performed. 
We report now an interesting, simple realization of this idea, that quickly emerged when studying the contorsion phase of symmetrized MAG theories. The cardinal symmetry is given by a relaxed form of projective symmetry, explicitly investigated in \cite{Percacci:2020ddy}, where the symmetry's degrees of freedom are lowered from four to just one $\xi_{\mu}(x) = \partial_{\mu}\Omega(x)$. 
%
\subsection{The model}
We now prove the existence of a healthy, radiatively stable, configuration in torsion-free MAG. The generic Lagrangian is, transcribing from \cite{Percacci:2020ddy}, of the form  
\bea \label{torsionFree}
&&S(g,A) =-\frac12\int d^4x\ \sqrt{-g}\,\Big[ -a_0 F+ F^{\mu\nu\rho\sigma} \big( c_1 F_{\mu\nu\rho\sigma} 
+ c_2 F_{\mu\nu\sigma\rho} 
+ c_3 F_{\rho\sigma\mu\nu} \big) +
\nn\w2
&&   
+ \,F^{(13)\mu\nu} \big(c_7 F^{(13)}_{\mu\nu} + c_8 F^{(13)}_{\nu\mu} \big)
+ F^{(14)\mu\nu} \big( c_9 F^{(14)}_{\mu\nu} 
+ c_{10} F^{(14)}_{\nu\mu}\big)  
+ F^{(14)\mu\nu}\big(c_{11} F^{(13)}_{\mu\nu}
+ c_{12} F^{(13)}_{\nu\mu} \big) +
\nn\w2
&& 
+\,c_{16}F^2 + Q^{\rho\mu\nu}\big( a_4 Q_{\rho\mu\nu} 
+ a_5 Q_{\nu\mu\rho}\big)  
+ a_6 Q^\mu Q_\mu + a_7 \tQ^\mu \tQ_\mu + a_8 Q^\mu \tQ_\mu
\Big]\ .
\eea
We constrain the 16 parameters\footnote{Upon moving in the contorsion phase we find, in the torsion-free case, that  $F^{\mu\nu\rho\sigma}(F_{\mu\nu\sigma\rho} + 2 F_{\rho\sigma\mu\nu} - 2 F_{\mu\rho\nu\sigma}) = 0$. In our Lagrangian we therefore omit the $F^{\mu\nu\rho\sigma} F_{\mu\rho\nu\sigma}$ term which is instead present in \cite{Percacci:2020ddy}.} with the non-accidental abelian symmetry 
\bea \label{AbelianProj}
&& \delta h_{\mu \nu} = 0 \, , \,\, \,\,\,\,\,\,\,\delta A_{\mu \,\, \nu}^{\,\,\,\rho}  =  g_{\mu \nu} \, \partial^{\rho} \Omega(x)\, ,
\eea 
finding the reduction 
\bea
&& a_7 = \tfrac{1}{25} \left(2 \, a_0 -10 \, a_4 - 3 \, a_5 + 4 \, a_6 \right)
\,, \nn \w2
&& a_8 = \tfrac{1}{10} \left(a_0 - 4 \,a_5 - 8 \,a_6 \right) \,, c_8 = - c_7 \,,
\eea
where the remaining $c_i$ are set to zero. In the contorsion phase we have a theory of the type (\ref{TableGray}) which, together with the Hilbert term $\sim {a_0 R}$, displays the following 
\bea \label{TheMasslessMode}
&& \mathcal L(k) = \tfrac{1}{2} (- a_0 - 2 a_4 - 3 a_5) k_{\alpha \beta \mu } k^{\alpha \mu \beta } + (- a_4 -  \tfrac{1}{2} a_5) k_{\alpha \mu \beta } k^{\alpha \mu \beta } \nn \w2 && + \tfrac{1}{25} (8 a_0 + 10 a_4 + 13 a_5 + 16 a_6) k^{\alpha }{}_{\alpha }{}^{\mu } k^{\beta }{}_{\mu \beta } + \tfrac{1}{50} (-2 a_0 + 10 a_4 + 3 a_5 - 4 a_6) k^{\alpha \mu }{}_{\alpha } k^{\beta }{}_{\mu \beta } \nn \w2 &&  + \tfrac{1}{50} (-7 a_0 + 10 a_4 + 23 a_5 - 64 a_6) k^{\alpha }{}_{\alpha }{}^{\mu } k_{\mu }{}^{\beta }{}_{\beta } -  \tfrac{1}{2} c_7 \nabla_{\beta }k_{\mu }{}^{\nu }{}_{\nu } \nabla^{\beta}k^{\alpha }{}_{\alpha }{}^{\mu }  + \tfrac{1}{2} c_7 \nabla^{\beta }k^{\alpha }{}_{\alpha }{}^{\mu } \nabla_{\mu} k_{\beta }{}^{\nu }{}_{\nu } \, . \nn \\
\eea
Notice that, while we referred to this type of system, in the contorsion phase, as \emph{decoupled}, such definition only concerns quadratic mixing. As expected, in (\ref{TheMasslessMode}) a $h-k$ interaction survives dictated by general covariance. It is easy to spot a kinetic term of the $U(1)$-symmetry type for the vector $K_{\mu} = k_{\mu }{}^{\nu }{}_{\nu }$ , it is less intuitive to realize that the quadratic non-derivative terms in $k_{\mu }{}^{\rho }{}_{\nu }$ do not introduce, as forced by the symmetry, any pole to the propagator. For this we carry a simple spectral analysis of the Lagrangian (\ref{TheMasslessMode}), ignoring the trivial results coming from studying the spectrum of the Einstein-Hilbert term. The symmetric $k_{\mu \,\,\,\nu}^{\,\,\,\rho} = k_{\nu \,\,\,\mu}^{\,\,\,\rho}$, that guarantees the absence of torsion, enjoys the decomposition
\begin{eqnarray}\label{TorsionSpin}
&k_{\mu \nu \rho} \supset & 3_1^- \oplus 2_1^+ \oplus 2_2^+ \oplus 2_1^- \oplus 1^+_1 \oplus 1^-_1 \oplus 1^-_2  \oplus 1^-_4 \oplus 1^-_5  \oplus  0^+_1 \oplus 0^+_2 \oplus 0^+_4 .
\end{eqnarray}
We then obtain the following verbose list of spin/parity matrices  
\bea %
&&a^{\left\lbrace 3,- \right\rbrace}_{1,1}  = - a_0 - 4 \left( a_4 + a_5\right) \,, \nn \w2 
&& a^{\left\lbrace 2,+ \right\rbrace}_{i,j} = 
\begin{pmatrix}
a_{1,1} & a_{1,2} \\
a_{2,1} & a_{2,2} 
\end{pmatrix}
=
\begin{pmatrix}
- a_0 - 4 \left( a_4 + a_5\right) & 0  \\
0 & \tfrac{1}{2}\left(a_0 - 2 a_4 + a_5\right) 
\end{pmatrix} \,, \nn \w2 
&&a^{\left\lbrace 2,- \right\rbrace}_{1,1}  = \frac{1}{2} \left( a_0 - 2 a_4 + a_5\right) \,, \nn \w2 
&&a^{\left\lbrace 1,+ \right\rbrace}_{1,1}  = \frac{1}{2} \left( a_0 - 2 a_4 + a_5\right) \,, \nn
\eea
\bea \label{TorsionSpinM}
&& a^{\left\lbrace 1,- \right\rbrace}_{i,j} = 
\begin{pmatrix}
a_{1,1} & a_{1,2} & a_{1,4}  & a_{1,5} \\
a_{2,1} & a_{2,2} & a_{2,4}  & a_{2,5} \\
a_{4,1} & a_{4,2} & a_{4,4}  & a_{4,5} \\
a_{5,1} & a_{5,2} & a_{5,4}  & a_{5,5}
\end{pmatrix}
= \nn \w2
&&\left( \begin{matrix}
 -\frac{4\,(2 a_0+5 a_4+2 a_5+9 a_6)}{15} -\frac{5 c_7
   q^2}{3} & -\frac{11 a_0+20 a_4-4 a_5+72 a_6+25
   c_7 q^2}{15 \sqrt{5}}  \\
 -\frac{11 a_0+20 a_4-4 a_5+72 a_6+25 c_7 q^2}{15
   \sqrt{5}} & \frac{(-19 a_0-130 a_4+41 a_5-288 a_6-50
   c_7 q^2)}{150}  \\
 \frac{7 a_0+40 a_4+52 a_5-36 a_6-25 c_7 q^2}{15
   \sqrt{5}} & \frac{(-11 a_0-20 a_4+4 a_5-72 a_6-25
   c_7 q^2)}{75}  \\
 -\frac{11 a_0+20 a_4-4 a_5+72 a_6+25 c_7 q^2}{15
   \sqrt{10}} & -\frac{47 a_0-10 a_4+17 a_5+144 a_6+25
   c_7 q^2}{75 \sqrt{2}} \\
\end{matrix} \right.
 \nn \w2 
 && \hspace{2.5cm} \left. \begin{matrix}
 \frac{7 a_0+40 a_4+52 a_5-36
   a_6-25 c_7 q^2}{15 \sqrt{5}} & -\frac{11 a_0+20 a_4-4
   a_5+72 a_6+25 c_7 q^2}{15 \sqrt{10}} \\
  \frac{(-11 a_0-20 a_4+4 a_5-72
   a_6-25 c_7 q^2)}{75}  & -\frac{47 a_0-10 a_4+17
   a_5+144 a_6+25 c_7 q^2}{75 \sqrt{2}} \\
  -\frac{4\,(17 a_0+65 a_4+62 a_5+9
   a_6)}{75} -\frac{c_7 q^2}{3} & -\frac{11 a_0+20 a_4-4
   a_5+72 a_6+25 c_7 q^2}{75 \sqrt{2}} \\
  -\frac{11 a_0+20 a_4-4 a_5+72
   a_6+25 c_7 q^2}{75 \sqrt{2}} & \frac{(14 a_0-70
   a_4+29 a_5-72 a_6)}{75} -\frac{c_7 q^2}{6} \\
\end{matrix} \right)
 \nn \w2 
&& a^{\left\lbrace 0,+ \right\rbrace}_{i,j} = 
\begin{pmatrix}
a_{3,3} & a_{3,5} & a_{3,6} \\
a_{5,3} & a_{5,5} & a_{5,6} \\
a_{6,3} & a_{6,5} & a_{6,6}
\end{pmatrix}
= \nn \w2
&& \begin{pmatrix}
 -\frac{6(3 a_0+10 a_4+8 a_5+6 a_6)}{25}  & \frac{11
   a_0+20 a_4-4 a_5+72 a_6}{25 \sqrt{2}} & \frac{(7
   a_0+40 a_4+52 a_5-36 a_6)}{25}  \\
 \frac{11 a_0+20 a_4-4 a_5+72 a_6}{25 \sqrt{2}} & \frac{(4 (-5 a_4+a_5-18 a_6)-11 a_0)}{25}
    & \frac{11 a_0+20
   a_4-4 a_5+72 a_6}{25 \sqrt{2}} \\
 \frac{(7 a_0+40 a_4+52 a_5-36 a_6)}{25}  & \frac{11
   a_0+20 a_4-4 a_5+72 a_6}{25 \sqrt{2}} & -\frac{6\,(3
   a_0+10 a_4+8 a_5+6 a_6)}{25}  \\
   \end{pmatrix} \,. \nn \\ 
\eea
From the cloudy structure of (\ref{TorsionSpinM}) it is possible to derive some simple, but probably not immediate, consequences. At first, we notice that only the sector $1^-$ can propagate a particle. Then, regardless the appearances, exploiting the determinants immediately reveals that such pole is massless and come with the gauge symmetry connected with the degeneracy of $a^{\left\lbrace 0,+ \right\rbrace}_{i,j}$, as typical in $U(1)$ symmetric models. The ultimate nature of this pole is exposed by the saturated propagator which, upon taking account for the symmetry, shows the simple structure 
\bea
\lim_{q^2 \rightarrow 0} \mathcal D_S = \frac{1}{q^2}X^{\dagger} \mathcal M \,X = \frac{1}{q^2} \left(\frac{1}{c_7}\right)\sum_{i = 1,2}|S_i|^2 \, .
\eea
We have then proved that the constrained system (\ref{TheMasslessMode}), stabilized by the symmetry (\ref{AbelianProj}), is free of ghosts without further tuning other than $c_7 > 0$. 
\subsection{The Stueckelberg sector of MAG}
The results of the previous section, while unsurprising,  might appear purely academical, given the phenomenological obstructions to an additional massless vector in the particles spectrum. Nevertheless, the experience with lower spin models, in particular the link between massless gauge modes and the Stueckelberg/Goldstone sectors, points to the necessary existence of a  supporting scalar counterpart. Exploring this hypothesis for higher-rank fields is a challenging task with an uncertain outcome. The inclusion of all the covariant interactions of extra scalars to eq.~(\ref{torsionFree}), in particular of new quadratic mixed terms, considerably modifies the starting theory.  While it is not difficult to recognize a possible extension of the symmetry and obtain a gauge-invariant mass term, the profusion of different possible contractions, typical of higher rank fields, endangers the stability requirement we sought at the beginning.  For our analysis to be coherent we must carefully supervise the possible introduction of extra poles and if radiatively unstable tuning is required. We entrust, for the spectral analysis of the larger space for fields of rank $3, 2,$ and $0$ the full set of projector operators that have been recently computed \cite{Percacci:2020ddy, Marzo:2021esg}. \\
We now start our investigation considering the simple extension with a single real scalar field $\phi(x)$ to the torsion-free MAG setup of eq.~(\ref{torsionFree})
\bea \label{torsionFreeStueck}
&&S(g,A,\phi) =\frac12\int d^4 x\ \sqrt{-g}\,\Big[\nabla_{\mu} \phi \nabla^{\mu} \phi + \left(s_0  + d_0 F \right)\phi^2 + \left(d_1 Q_{\rho \,\,\,\mu}^{\,\,\,\mu} + d_2 Q^{\mu}_{\,\,\,\rho \mu} \right) \nabla^{\rho} \phi 
\Big]\ . \nn \\
\eea
We the test the constrains over $S(g,A) + S(g,A,\phi)$ of the familiar Stueckelberg ansatz
\bea \label{AbelianProjStu}
&& \delta h_{\mu \nu} = 0 \, , \,\, \,\,\,\,\,\,\,\delta A_{\mu \,\, \nu}^{\,\,\,\rho}  =  g_{\mu \nu} \, \partial^{\rho} \Omega(x)\, \, , \,\, \,\,\,\,\,\,\,\delta \phi = f\,  \Omega(x)\, ,
\eea 
and find non-trivial solutions for $f = - d_1 - \tfrac{5}{2} d_2$ and
\bea
&& a_7 = \tfrac{1}{25} \left(2 \, a_0 -10 \, a_4 - 3 \, a_5 + 4 \, a_6 +  d_1^2\right) -  \frac{d_2^2}{4} 
\,, \nn \w2
&& a_8 = \tfrac{1}{10} \left(a_0 - 4 \,a_5 - 8 \,a_6 - 2\, d_1^2 - 5 \,d_1 d_2 \right) \,, \nn \w2
&& c_8 = - c_7 \,\,,\,\text{other\,\,}c_{i} = 0\,  .
\eea
At the price of two extra parameters we extended the shift symmetry (\ref{AbelianProj}) to coherently include an extra scalar field. In the contorsion phase we arrive at the following addition to the Einstein-Hilbert term
\bea
&&\mathcal L (k, \phi)= \tfrac{1}{2} (- a_0 - 2 a_4 - 3 a_5) k_{\alpha \beta \mu } \
k^{\alpha \mu \beta } + (- a_4 -  \tfrac{1}{2} a_5) k_{\alpha \mu \
\beta } k^{\alpha \mu \beta } + \nn \w2 &&  + \frac{R_1}{200}  k^{\alpha \mu \
}{}_{\alpha } k^{\beta }{}_{\mu \beta } + \frac{R_2}{100} k^{\alpha }{}_{\alpha 
}{}^{\mu } k^{\beta }{}_{\mu \beta } + \frac{R_3}{200} k^{\alpha \
}{}_{\alpha }{}^{\mu } k_{\mu }{}^{\beta }{}_{\beta } + \nn \w2 && + (d_1 + \
\tfrac{1}{2} d_2) k_{\alpha }{}^{\mu }{}_{\mu } \nabla^{\alpha \
}\phi + \tfrac{1}{2} d_2 k^{\mu }{}_{\alpha \mu } \nabla^{\alpha }\phi \
+ \tfrac{1}{2} \nabla_{\alpha }\phi \nabla^{\alpha }\phi + \nn \w2 && -  \
\tfrac{1}{2} c_7 \nabla_{\beta }k_{\mu }{}^{\nu }{}_{\nu } \
\nabla^{\beta }k^{\alpha }{}_{\alpha }{}^{\mu }  + \tfrac{1}{2} c_7 \
\nabla^{\beta }k^{\alpha }{}_{\alpha }{}^{\mu } \nabla_{\mu \
}k_{\beta }{}^{\nu }{}_{\nu } \,,\nn\\
\eea
where we introduced the abbreviations
\bea \label{abbrv}
&&R_1 = (-8 a_0 + 40 a_4 + 12 a_5 - 16 a_6 - 4 d_1^2 + 25 d_2^2) \, , \nn \w2
&&R_2 = (32 a_0 + 40 a_4 + 52 a_5 + 64 a_6 + 16 d_1^2 + 50 d_1 d_2 + 25 d_2^2) \, , \nn \w2
&&R_3 = (-28 a_0 + 40 a_4 + 92 a_5 - 256 a_6 + 36 d_1^2 + 100 d_1 d_2 + 25 d_2^2) \, . 
\eea
Now, by processing this Lagrangian through our enlarged set of projector operators (and adding the sector $0^+_8$ carried by $\phi$ to (\ref{TorsionSpin})) we can show that the quadratic terms do generate a unique massive pole while preserving the defining abelian symmetry (\ref{AbelianProjStu}). The only spin/parity sectors different from (\ref{TorsionSpinM}) turn to be the $1^-$ 
\bea \label{TorsionSpinM2a}
&& a^{\left\lbrace 1,- \right\rbrace}_{i,j} = 
\begin{pmatrix}
a_{1,1} & a_{1,2} & a_{1,4}  & a_{1,5} \\
a_{2,1} & a_{2,2} & a_{2,4}  & a_{2,5} \\
a_{4,1} & a_{4,2} & a_{4,4}  & a_{4,5} \\
a_{5,1} & a_{5,2} & a_{5,4}  & a_{5,5}
\end{pmatrix}
= \nn \w2 \nn \w2
&&=\left( \begin{matrix}
 \tfrac{-60 \,a_0 - 240 \,a_4 -240 \,a_5 - 100 \,c_7 \,q^2 + R_1 + 2 \,R_2 + R_3}{60} & - \tfrac{100 \,c_7 \,q^2 + 2 \,R_1 + R_2 - R_3}{60 \sqrt{5}}  \\
- \tfrac{100 \,c_7 \,q^2 + 2 \,R_1 + R_2 - R_3}{60 \sqrt{5}} & \tfrac{150 \,a_0 - 300 \,a_4 + 150 \,a_5 -100 \,c_7 \,q^2 + 4 \,R_1 - 4 \,R_2 + R_3}{300}  \\
 - \tfrac{100 \,c_7 \,q^2 +  R_1 + 2 \,R_2 + R_3}{60 \sqrt{5}} & \tfrac{-100 \,c_7 \,q^2 - 2 \,R_1 - R_2 + R_3}{300} \\
- \frac{100 \,c_7 \,q^2 + 2 \,R_1 + R_2 - R_3}{60 \sqrt{5}}  & - \tfrac{100 \,c_7 \,q^2 + 4 \,R_1 - 4 \,R_2 + R_3}{300 \sqrt{2}}  \\
\end{matrix} \right.
 \nn \w2 
 && \hspace{2cm} \left. \begin{matrix}
 \tfrac{-100 \,c_7 \,q^2 +  R_1 + 2 \,R_2 + R_3}{60 \sqrt{5}}  & - \tfrac{100 \,c_7 \,q^2 + 2 \,R_1 + R_2 - R_3}{60 \sqrt{10}}  \\
\tfrac{- 100 \,c_7 \,q^2 - 2 \,R_1 - R_2 + R_3}{300} & \tfrac{-100 \,c_7\, q^2 + 4 \,R_1 - 4 \,R_2 + R_3}{300 \sqrt{2}}  \\
 \tfrac{- 300 \,a_0 - 1200 \,a_4 - 1200 \,a_5 - 100 \,c_7 \,q^2 + R_1 + 2 \,R_2 + R_3}{300} & - \tfrac{100 \,c_7 \,q^2 + 2 \,R_1 + R_2 - R_3}{300 \sqrt{2}}  \\
- \tfrac{100 \,c_7 \,q^2 + 2 \,R_1 + R_2 - R_3}{300 \sqrt{2}} & \tfrac{4\left(75 \,a_0 - 15 \,a_4 + 75 \,a_5 - 25 \,c_7 q^2 + R_1 - R_2\right) + R_3}{600} \\
\end{matrix} \right)
 \nn \w2 
\eea
and the $0^+$ one
\bea
&& a^{\left\lbrace 0,+ \right\rbrace}_{i,j} = 
\begin{pmatrix}
a_{0,0} & a_{0,3} & a_{0,5} & a_{0,6} \\
a_{3,0} & a_{3,3} & a_{3,5} & a_{3,6} \\
a_{5,0} & a_{5,3} & a_{5,5} & a_{5,6} \\
a_{6,0} & a_{6,3} & a_{6,5} & a_{6,6}
\end{pmatrix}
= \nn \w2 \nn \w2  
&&=\left( \begin{matrix}
 \tfrac{-100 \,a_0 - 400 \,a_4 - 400 \,a_5  + R_1 + 2 \,R_2 + R_3}{60} & \tfrac{2 \,R_1 + R_2 - R_3}{100 \sqrt{2}}  \\
\tfrac{2 \,R_1 + R_2 - R_3}{100 \sqrt{2}} & \tfrac{4 \left(25 \,a_0 - 50 \,a_4 + 25 \,a_5 + R_1 - R_2\right) + R_3}{200}  \\
 \tfrac{R_1 + 2 \,R_2 + R_3}{100} & \tfrac{2 \,R_1 + R_2 - R_3}{100 \sqrt{2}} \\
 - i \left( d_1 + d_2\right) \sqrt{q^2}  & \tfrac{i \left(2 d_1 - d_2\right)\sqrt{q^2}}{2 \sqrt{2}}  \\
\end{matrix} \right.
 \nn \w2 
 && \hspace{4cm} \left. \begin{matrix}
 \tfrac{R_1 + 2 \,R_2 + R_3}{100}  & i \left(d_1 + d_2 \right) \sqrt{q^2}  \\
\tfrac{2 \,R_1 + R_2 - R_3}{100 \sqrt{2}} & - \tfrac{i \left(2 \,d_1 - d_2 \right)\sqrt{q^2}}{2 \sqrt{2}}  \\
 \tfrac{- 100 \,a_0 - 400 \,a_4 - 400 \,a_5  + R_1 + 2 \,R_2 + R_3}{100} & i \left(d_1 + d_2\right) \sqrt{q^2}  \\
-i \left(d_1 + d_2\right)\sqrt{q^2} & q^2 \\
\end{matrix} \right) \, .
 \nn \w2 
\eea
The degeneracy of the $a^{\left\lbrace 0,+ \right\rbrace}_{i,j}$ shows that the model still preserved the extended abelian gauge invariance of (\ref{AbelianProjStu}), meanwhile from the $1^-$ sector, we find the unique \emph{massive} pole 
\bea
m^2_{1^-} = \frac{\left(2 \,d_1 + 5 \,d_2\right)^2\,\left(50 \left(2 \,d_1 - d_2\right)\left(d_1 + d_2\right) + 2 \,R_1 + R_2 - R_3\right)}{4 \,c_7 \left(50 \left(d_1 - 2 \,d_2\right)\left(2 \,d_1 + 5 \,d_2\right)+ 2\, R_1 + R_2 - R_3\right)} \,,
\eea
where, for aesthetic benefit,  we traded the parameters $a_{4,5,6}$ for the triplet $R_i$. Taking into account gauge invariance via the constrained source (\ref{eq12}), we find 
\bea
\lim_{q^2 \rightarrow m^2_{1^-}} \mathcal D_S = \frac{X^{\dagger} \mathcal M \,X}{q^2- m^2_{1^-}} = \frac{Res_{1^-}}{q^2 -  m^2_{1^-}} \sum_{i = 1,3}|S_i|^2 \, ,
\eea  
with the residue $Res_{1^-}$ given by  
\bea
Res_{1^-} = \frac{4}{c_7}\left[ 1 + \frac{75\, d_2 \,(2\, d_1 + 5\, d_2)(50 \,(3 \,d_1 - 2 \,d_2)(2\, d_1 + 5\, d_2) + 3(2\, R_1 + R_2 - R_3))}{(50 (d_1 - 2 \,d_2)(2 \,d_1 + 5 \,d_2) + 2 \,R_1 + R_2 - R_3)^2} \right] \,. \nn \\
\eea
We can now require simultaneous positivity of the pole and the residue. For real couplings, and with $a_0 < 0$ from the graviton $2^+$ sector, different possible solutions are produced. As anticipated, a subset of these are equalities that re-introduce the tuning we planned to avoid, being unrelated to the symmetry of the theory. We dispose of these. The success of our quest is instead measured by the existence of relations which can be more naturally accommodated without tuning to zero the starting operators. 
Among such solutions we find, together with $a_0 <0$ and $c_7 > 0$, the following alternatives
\bea \label{finalC}
I) &&A \neq 0 \,, B >0 \nn \w2
&& \left[(d_1 \neq 0 \, , 2\, d_1 + 5\, d_2 \neq 0) \lor \left(d_1 < 0 \, , d_2 \leq 0\right) \lor \left(d_1 > 0 \, , d_2 \geq 0\right) \right] \nn \w2
II) &&A < 0\nn \w2 && 
\left[(d_1 \neq 0 \, , 2\, d_1 + 5\, d_2 \neq 0) \lor \left(d_1 < 0 \, , d_2 \leq 0\right) \lor \left(d_1 > 0 \, , d_2 \geq 0\right) \right] \nn \w2
III) && B > 0\nn \w2 && 
\left[\left(d_1 < 0, d_2 < 0 \right) \lor \left(d_1 < 0,\, 2 d_1 + 5 d_2 > 0 \right) \lor \left(d_1 < 0, d_2 >0, 2 d_1 + 5 d_2 < 0 \right) \right] \nn \w2
IV) && B > 0\nn \w2 && 
\left[\left(d_1 > 0, d_2 > 0 \right) \lor \left(d_1 > 0, 2 d_1 + 5 d_2 < 0 \right) \lor \left(d_1 > 0, d_2 < 0 \, , 2 d_1 + 5 d_2 > 0 \right) \right] \, , \nn \\
\eea
with the abbreviations 
\bea
&&A = 11 \,a_0 + 20 \, a_4 + 72\, a_6 + 18\, d_1^2 - 4\, a_5 \, ,\nn \w2
&& B = 22 \,a_0 + 40 \, a_4 + 36(4\, a_6 + d_1^2)- 8\, a_5 - 225 \,d_2^2 \, .\nn
\eea
The constrains resemble the consistency conditions met in vacuum stability analyses. 
They select regions in the parameter space cruised by the renormalization group flow. Two scenarios are then possible. Found a point for which any of (\ref{finalC}) is valid, the renormalization group trajectory will asymptotically remain in the corresponding region. More likely, a cut-off energy will signal the outward crossing of the safe region marking the range of validity of our description. 
\section{Conclusions}
In theories with a large volume of couplings, the search for unitary and causal propagation demands precise cancellations to occur among unrelated parameters. In turn, these are reduction statements that connect different operators in the Lagrangian. When these constraints generate an accidental symmetry, only realized at the quadratic level, no hope is given that the loop corrections will preserve it. Interactions will bring back the ghost we hunted away once the gauge-fixed propagator will couple to a non-conserved source. Even when no accidental symmetries are generated, radiative corrections will harass the propagator and ask for continued tuning to cancel them away.  There are no immediate, consistency arguments against the latter case. It is nevertheless interesting to push the idea that this scenario should be avoided, following what is realized in the lower spin case.  
In this work, we promote the principle that the search for ghosts- and tachyons-free regions, in the parameter space of MAG, must be supported by the requirement of stability against radiative corrections. This challenging prerequisite greatly narrows the space of acceptable solutions and asks for the support of opportune non-accidental symmetries. By a guided exploration of the possible symmetries and their effect over the Lagrangian, we proved that  MAG inhabits a massive vector state in a sector stabilized by an additional abelian symmetry. The proof demanded the introduction of an extra scalar providing the missing Stueckelberg-type degree of freedom, and dedicated spectral analysis of the mixed terms for fields of rank $3, 2$ and $0$. Many future directions of our approach can be identified. We stress, in particular, the exploration of the interplay between symmetries and quadratic curvature terms in the so-called \emph{coupled} scenario. From the phenomenological side, thorough profiling of the massive vector is in order. This is particularly promising if we consider the many applications that rely on a massive abelian vector to tackle long-standing anomalies \cite{DeFelice:2016yws,Heisenberg:2020xak,Arcadi:2020jqf}. In our opinion, the principle of radiative stability that supports our massive vector selects a more solid path towards the discovery of new degrees of freedom in MAG theories. 
  
\subsection*{Acknowledgments}
We thank Marco Piva for precious suggestions. This work was supported by the Estonian Research Council grant PRG803
and by the EU through the European Regional Development Fund
CoE program TK133 ``The Dark Side of the Universe." 
 \pagebreak


\end{document}